\documentclass[fontsize=11pt]{article}

\usepackage{
  amsmath,
  amsthm,
  amssymb,
}

\usepackage[hidelinks]{hyperref}
\usepackage{cleveref}
\usepackage{amsfonts}
\usepackage{mathrsfs}
\usepackage{graphicx}

\title{A Survey On Secure Machine Learning}

\author{
   Taobo Liao \quad \textbar \quad Taoran Li \quad \textbar \quad Prathamesh Nadkarni \\
  \texttt{taobol2\textbar{}taoranl2\textbar{}pyn2} \\
  \texttt{@illinois.edu}\\
  University of Illinois Urbana-Champaign\\
  Urbana, IL 61801 \\
}
  
\date{April 22, 2024}

\begin{document}
\maketitle

\begin{abstract}
In this survey, we will explore the interaction between secure multiparty computation and the area of machine learning. Recent advances in secure multiparty computation (MPC) have significantly improved its applicability in the realm of machine learning (ML), offering robust solutions for privacy-preserving collaborative learning. This review explores key contributions that leverage MPC to enable multiple parties to engage in ML tasks without compromising the privacy of their data. The integration of MPC with ML frameworks facilitates the training and evaluation of models on combined datasets from various sources, ensuring that sensitive information remains encrypted throughout the process. Innovations such as specialized software frameworks and domain-specific languages streamline the adoption of MPC in ML, optimizing performance and broadening its usage. These frameworks address both semi-honest and malicious threat models, incorporating features such as automated optimizations and cryptographic auditing to ensure compliance and data integrity. The collective insights from these studies highlight MPC's potential in fostering collaborative yet confidential data analysis, marking a significant stride towards the realization of secure and efficient computational solutions in privacy-sensitive industries. This paper investigates a spectrum of SecureML libraries that includes cryptographic protocols, federated learning frameworks, and privacy-preserving algorithms. By surveying the existing literature, this paper aims to examine the efficacy of these libraries in preserving data privacy, ensuring model confidentiality, and fortifying ML systems against adversarial attacks. Additionally, the study explores an innovative application domain for SecureML techniques: the integration of these methodologies in gaming environments utilizing ML.
\end{abstract}

\section{Introduction}\label{sec:intro}

Secure multiparty computation (MPC) has emerged as a pivotal technology enabling privacy-preserving machine learning (ML) across various domains. This review explores three significant contributions to the field:
\begin{enumerate}
    \item Secure multi-party computation (MPC) is a transformative cryptographic protocol that enables multiple parties to collaboratively engage in machine learning tasks without compromising the privacy of their individual data inputs. This paper \cite{cryptoeprint:2017/396} delves into the application of MPC in machine learning, showcasing recent advancements and frameworks that facilitate privacy-preserving computations in a variety of applications. We will introduce this part in \cref{sec:secureml}.
    
    \item The CRYPTEN \cite{NEURIPS2021_27545182} framework from Facebook AI Research exemplifies a significant stride in making MPC protocols accessible to machine learning practitioners who lack deep cryptographic expertise. CRYPTEN integrates seamlessly with popular machine learning frameworks, using secure MPC primitives to support private model evaluation and training. The framework is designed to work under a semi-honest threat model and utilizes GPU acceleration to enhance performance, demonstrating its utility with state-of-the-art models for text classification, speech recognition, and image classification. It aims to lower the barrier to adoption of MPC in machine learning by providing a user-friendly interface that mimics conventional machine learning APIs while ensuring secure computations. We will introduce this part in \cref{sec:crypten}.

    \item Another pivotal development is Cerebro \cite{272147}, a comprehensive platform for MPC-based collaborative learning detailed in the USENIX Security 2021 proceedings. Cerebro addresses both the generality-performance tradeoff and the privacy-transparency challenge inherent in collaborative learning environments. It features a Python-like domain-specific language that allows users to write arbitrary learning programs, which are then optimized by an advanced compiler that selects the most efficient MPC protocols for execution. Additionally, Cerebro introduces innovative mechanisms such as compute policies and cryptographic auditing, enabling organizations to control the release of computed results and to audit the computation process without revealing private data. We will introduce this part in \cref{sec:cerebro}.

\end{enumerate}

Both platforms, CRYPTEN and Cerebro, are pivotal in advancing how sensitive data is utilized securely in machine learning. By facilitating collaborative learning without data exposure, these MPC implementations not only enhance data security, but also enable the realization of more sophisticated machine learning models. This review underscores the potential of MPC to revolutionize fields that require stringent data security measures, paving the way for widespread adoption in healthcare, finance, and beyond.

Central to the advancement of SecureML are specialized libraries designed to implement privacy-preserving techniques and cryptographic protocols seamlessly within ML workflows. In this paper, we focus on five prominent libraries: JustGarble, TFEncrypted, Crypten, LibOTe, and OTExtension. These libraries, while diverse in their approaches, share common objectives and offer essential functionalities for enhancing privacy and security in ML applications. For this part, refer to \cref{sec:libraries}. We will also introduce the application of secure ML in the area of game development in \cref{sec:games}. 

\section{Secure ML: A System for Scalable Privacy-Preserving Machine Learning}\label{sec:secureml}
\subsection{Paper Abstract}
SecureML presents a new and efficient approach to privacy-preserving machine learning, tackling the challenge of scalability in securely training models such as linear regression, logistic regression, and neural networks using stochastic gradient descent. The protocol operates under a two-server model where data owners distribute encrypted data across two non-colluding servers. These servers perform the computation without learning anything about the underlying data, using secure two-party computation (2PC).
\subsection{Preliminaries}
\subsubsection{Machine Learning}
\paragraph{Linear regression}
Given n training data samples $x_i$ each containing d features and the corresponding output labels $y_i$, regression is a statistal process to learn a function such that $g(x_i) \approx y_i $.
\paragraph{Stochastic gradient descent (SGD)} SGD is an eﬀective approximation algorithm for approaching a local minimum of a function, step by step. The SGD algorithm works as follows: $w$ is initialized as a vector of random values or all $0$s. In each iteration, a sample ($x_i$, $y_i$) is selected randomly and a coeﬃcient $w_j$ is updated as
\[ w_j := w_j - \alpha \frac{\partial C_i(w)}{\partial w_j} \]
\paragraph{Mini-batch} In practice, instead of selecting one sample of data per iteration, a small batch of samples are selected randomly and $w$ is updated by averaging the partial derivatives of all samples on the current $w$.
\paragraph{Learning rate adjustment} If the learning rate $\alpha$ is too large, the result of SGD may diverge from the minimum. Therefore, a testing dataset is used to test the accuracy of current $W$ and adjust the leaning rate $\alpha$.
\paragraph{Termination} If the accuracy is lower than a specific threshold, we could terminate the learning process and output $w$.
\paragraph{Logistic Regression} Logistic Regression is used to classify the dataset into two classes, the output label $y$ is binary. Compared to linear regression, the logistic regression have an extra layer on which an activation function is applied on the result.
\paragraph{Neural Network} Neural networks are a generalization of regression to learn more complicated relationships between high dimensional input and output data.
\subsubsection{Secure Computation}
\paragraph{Oblivious Transfer} In an oblivious transfer protocol, a sender $S$ has two inputs $x_0$ and $x_1$, and a receiver $R$ has a selection bit $b$ and wants to obtain $x_b$ without learning anything else or revealing $b$ to $S$, which is denoted as 
\[ (\perp;x_b) \leftarrow OT(x_0,x_1;b)\]
The paper also use correlated OT extension \cite{cryptoeprint:2013/552}. In the COT, the sender $S$'s input is related, which is $x_0 = f(x_1)$.
\paragraph{Garbled Circuit 2PC} Garbled circuit could be used to securely compute any function by two parties. The paper require the garbling scheme to satisfy the standard security properties formalized in \cite{cryptoeprint:2012/265}.
\paragraph{Secret Sharing and Multiplication Triplets} The paper use secret share to perform secure computation, which contains additively share $Shr^A(\cdot)$, reconstruction $Rec^A(\cdot,\cdot)$ and multiply share $Mul^A(\cdot,\cdot)$. 
I want to go detail about the multiply share $Mul^A(\cdot,\cdot)$, which take advantage of Beaver's pre-computed multiplication triplet technique. The two parties have shared $\langle u \rangle$,$\langle v \rangle$,$\langle z \rangle$, where $u,v \xleftarrow{\$} \mathbb{Z}_{2^\mathscr{l} }$ and $z = uv \, mod \, 2^\mathscr{l}$. Then $P_i$ locally computes $\langle e \rangle = \langle a \rangle - \langle u \rangle$ and $\langle f \rangle = \langle b \rangle - \langle v \rangle$. Both parties run $Rec(\langle e_0 \rangle,\langle e_1 \rangle)$ and $Rec(\langle f_0 \rangle,\langle f_1 \rangle)$ and $P_i$ let $\langle c_i \rangle = -i \cdot e \cdot f + f \cdot \langle a \rangle_i + e \cdot \langle b \rangle_i + \langle z \rangle_i$. Note that Boolean sharing can be seen as additive sharing in $\mathbb{Z}_{2^\mathscr{l}}$.
 
\subsection{Secure Model}
Let $C_1,...,C_m$ denote the clients and $S_0,S_1$ denote two non-colluding servers. We assume a semi-honest adversary $A$ who can corrupt any subset of the clients and at most one of the two servers. The ideal functionality is described in figure \ref{fig:functionality}.
\begin{figure}[t]\centering
\framebox{
\begin{minipage}{0.95\linewidth}
{\sc Parameters:}  
Clients $C_1,...,C_m$ and servers $S_0$,$S_1$. \\
{\sc Uploading Data:}
On input $x_i$ from $C_i$, store $x_i$ internally.\\
{\sc Computation:}
On input $f$ from $S_0$ or $S_1$, compute $(y1,...,y_m) = f(x_1,...,x_m)$ and send $y_i$ to $C_i$. This step can be repeated multiple times with different functions.
\end{minipage}
}
\caption{Ideal functionality}
\label{fig:functionality}
\end{figure}

\subsection{Privacy Preserving Linear Regression}
\subsubsection{Vectorization}
In order to benefit from the mini-batch and vectorization, the paper generalizes the addition and multiplication operations on share values to shared matrices.
\subsubsection{Truncation}
In order to improve the efficiency, the authors use a technique called truncation, which simply truncate the last half part of the decimal part of the recovered number and keep the front half part. The authors proved that this technique has a small influence on the accuracy of the final result. The theorem is described in Figure \ref{fig:truncation}.

\begin{figure}[t]\centering
\framebox{
\begin{minipage}{0.95\linewidth}
{\sc Theorem 1:}  
In field $\mathbb{Z}_{2^\mathscr{l} }$, let $x \in \lbrack 0,2^{l_x} \rbrack \cup \lbrack 2^l - 2^{l_x}, 2^l \rbrack$, where $l>l_x+1$ and given shares $\langle x \rangle_0$, $\langle x \rangle_1$ of $x$, let $\langle \lfloor x \rfloor \rangle_0 = \lfloor \langle  x  \rangle_0  \rfloor$ and $\langle \lfloor x \rfloor \rangle_1 = 2^l - \lfloor 2^l - \langle  x  \rangle_1  \rfloor$. Then with probability $1-2^{l_x +1-l}$, $Rec(\langle \lfloor x \rfloor \rangle_0, \langle \lfloor x \rfloor \rangle_1) \in \{ \lfloor x \rfloor -1, \lfloor x \rfloor, \lfloor x \rfloor +1 \}$, where $\lfloor \cdot \rfloor$ denotes truncation by $l_D \leq l_x$ bits. 
\end{minipage}
}
\caption{Truncation Theorem}
\label{fig:truncation}
\end{figure}

\subsubsection{The online phase of privacy preserving linear regression}

\begin{figure}
\centering
\includegraphics[width=1\linewidth]{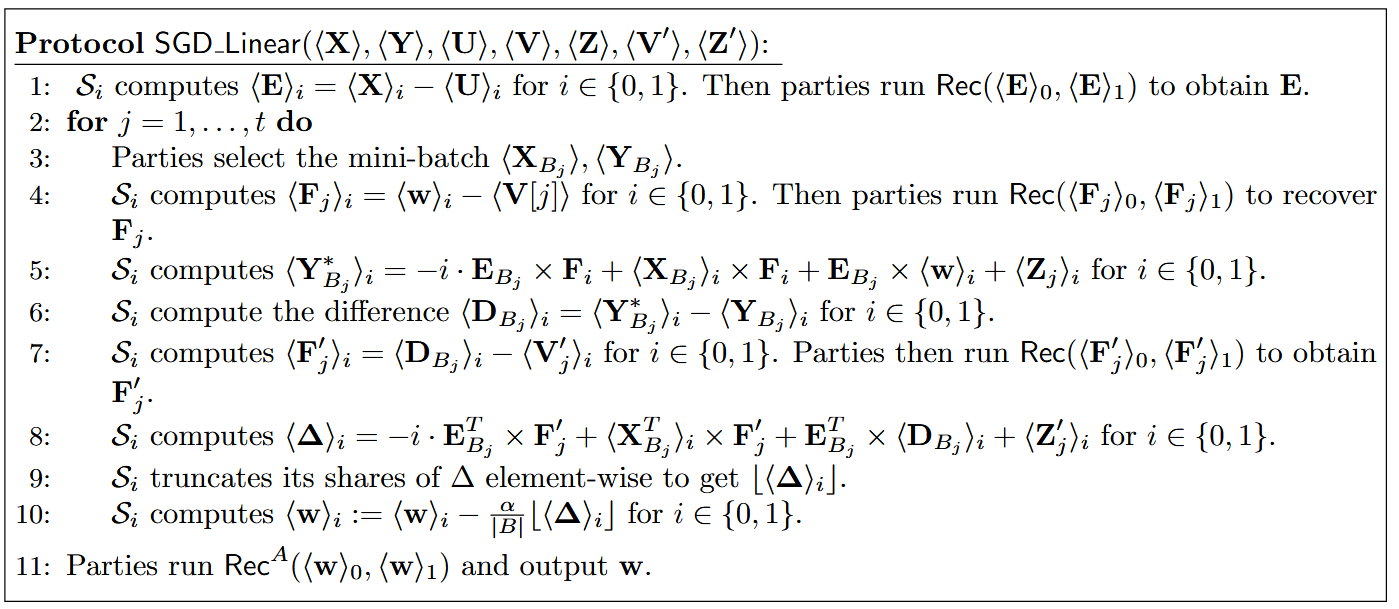}
\caption{\label{fig:f3}The online phase of privacy preserving linear regression}
\end{figure}
The online phase of privacy preserving linear regression is described in Figure \ref{fig:f3}. It assumed that the data-independent shared matrices $\langle U \rangle$, $\langle V \rangle$, $\langle Z \rangle$, $\langle V' \rangle$, $\langle Z' \rangle$ were already generated in the offline phase. The protocol also requires multiplying the coefficient vector by $\frac{\alpha}{\vert B \vert}$ in each iteration, and $\frac{\alpha}{\vert B \vert}$ is set to be a power of 2, i.e. $\frac{\alpha}{\vert B \vert} = 2^{-k}$. Then the multiplication with $\frac{\alpha}{\vert B \vert}$ can be replaced by having the parties truncate $k$ additional bits from their shares of the coefficients. 
\subsubsection{The offline phase of privacy preserving linear regression}
In the offline phase, the authors need to compute $C = A \times B  = \langle A \rangle_0 \times \langle B \rangle_0 + \langle A \rangle_0 \times \langle B \rangle_1 + \langle A \rangle_1 \times \langle B \rangle_0 + \langle A \rangle_1 \times \langle B \rangle_1 $. It is sufficient to compute $\langle \langle A \rangle_0 \times \langle B \rangle_1 \rangle $ and $ \langle \langle A \rangle_1 \times \langle B \rangle_0 \rangle$ because the other two terms can be computed locally.

\begin{figure}
\centering
\includegraphics[width=1\linewidth]{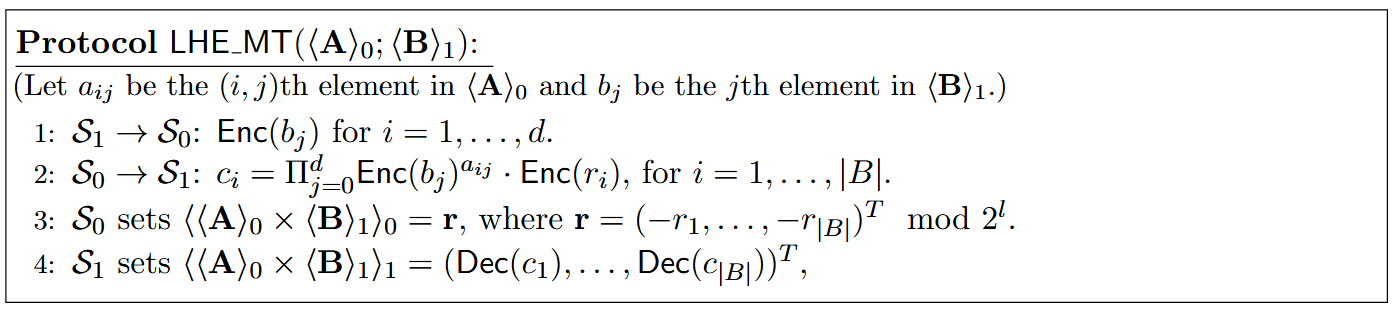}
\caption{\label{fig:f4}The oﬄine protocol based on linearly homomorphic encryption}
\end{figure}

\paragraph{LHE-based generation} The LHE-based generation involves the technique named linearly homomorphic encryption (LSE). The protocol can be found in Figure \ref{fig:f4}.

\paragraph{OT-base generation} 
The shares of the product $\langle A \rangle_0 \times \langle B \rangle_1$ could also be computed using OTs. The full protocol is a bit complex so that I will not go detail in this protocol. The full protocol is described detailed in \cite{cryptoeprint:2017/396}.

\subsection{Privacy Preserving Logistic Regression}
\subsubsection{Secure computation friendly activation function}
\begin{figure}
\centering
\includegraphics[width=0.5\linewidth]{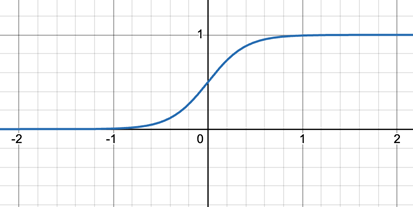}
\caption{\label{fig:f5}Sigmoid function}
\end{figure}

\begin{figure}
\centering
\includegraphics[width=0.5\linewidth]{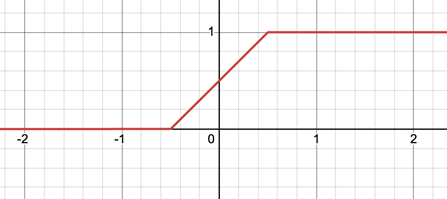}
\caption{\label{fig:f6}Secure computation friendly activation function}
\end{figure}

Because of the logistic activation Sigmoid function, described in Equation \ref{eq:e1} and Figure \ref{fig:f5}, cannot be computed by multiplication and additive share, the author propose a new activation function that can be eﬃciently computed using secure computation techniques. The function is described in Equation \ref{eq:e2} and Figure \ref{fig:f6}.
\begin{equation}
f(x)=\frac{1}{1+e^{-x}}
\label{eq:e1}
\end{equation}

\begin{equation}
f(x)=\left\{
\begin{array}{rcl}
0 & & {x< -\frac{1}{2}}\\
x + \frac{1}{2} & & {-\frac{1}{2} \leq x \leq \frac{1}{2}}\\
1 & & {x \geq \frac{1}{2}}
\end{array} \right.
\label{eq:e2}
\end{equation}

\subsubsection{Privacy preserving logistic regression protocol}
\begin{figure}
\centering
\includegraphics[width=1\linewidth]{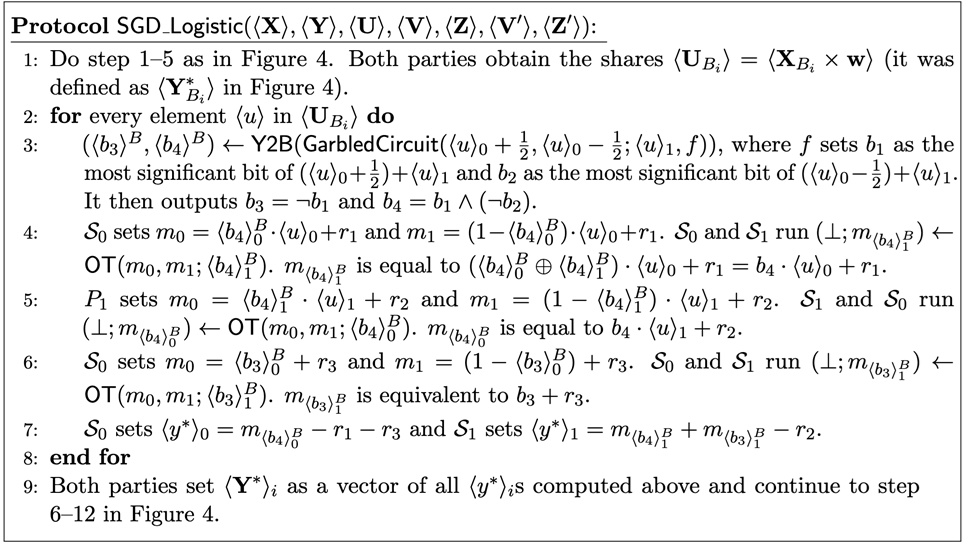}
\caption{\label{fig:f7}Privacy preserving logistic regression protocol}
\end{figure}

Firstly, we need to define two variables as described in Equation \ref{eq:e3} and Equation \ref{eq:e4}. Use these two variables, we could update the activation function in Equation \ref{eq:e2} to 
\[ f(u) = (\neg b_2) + (b_2 \wedge (\neg b_1))u \]
Use thie activation function, the full protocol of Privacy preserving logistic regression is described in Figure \ref{fig:f7}.
\begin{equation}
b_1=\left\{
\begin{array}{rcl}
0 & & {u + \frac{1}{2} \geq 0}\\
1 & & {u + \frac{1}{2} < 0}
\end{array} \right.
\label{eq:e3}
\end{equation}

\begin{equation}
b_2=\left\{
\begin{array}{rcl}
0 & & {u - \frac{1}{2} \geq 0}\\
1 & & {u - \frac{1}{2} < 0}
\end{array} \right.
\label{eq:e4}
\end{equation}

\section{CRYPTEN: Secure Multi-Party Computation Meets Machine Learning} \label{sec:crypten}
\subsection{Introduction}
CRYPTEN introduces a novel software framework designed to bridge the gap between secure multi-party computation (MPC) and machine learning. Developed by Facebook AI Research, it aims to make secure MPC techniques accessible and practical for machine learning researchers and engineers who may not have a deep background in cryptography.
\subsection{Framework Design}
CRYPTEN is structured around a PyTorch-like API, offering familiar machine learning abstractions, such as tensors, automatic differentiation, and neural networks, all designed to operate within a secure MPC environment. This design allows machine learning practitioners to leverage secure MPC methods without needing to understand the underlying cryptographic details.
\subsection{GPU Support and Scalability}
A distinguishing feature of CRYPTEN is its support for GPU computations, which significantly enhances the performance of MPC operations, allowing for more complex and larger scale machine learning tasks to be conducted securely and efficiently.
\subsection{Flexible Use Case Applications}
The framework is demonstrated to be capable of supporting a wide range of machine learning tasks from text classification to speech recognition and image classification, providing a robust tool for privacy-preserving machine learning.

\section{Cerebo - A Real-World implementation} \label{sec:cerebro}

\subsection{Platform Overview}
  \subsubsection{Threat model}
      After discussing the underlying building-blocks of Secure-ML, we will go through a recent platform that allows real-world users to perform those ideal functionalities with no prior knowledge about MPC or Secure-ML. The platform named Cerebro supports $P$-parties to perform a single learning task with their secret datasets. It also allows users to choose between two threat modes: semi-honest and malicious settings. Both settings are defined in the context of cryptography. In other words, the model can only prevent information leakage of users' datasets when $n-1$ parties are corrupted and form a semi-honest adversary or there are parties deviate from the protocol. In the high level, how Cerebo achieves both is by pre-defined sub-protocols and inconsistent data which will appear when misbehavior happening. However, in machine learning world, instead of the protocol, the trained model could also be easily attacked by carefully constructed dataset. An interesting example is that a skin cancer detection model\footnote{Problematic cancer detection model \url{https://www.sciencedirect.com/science/article/pii/S0022202X18322930}}  had mistakenly thought every image that contained ruler marking was indicative of melanoma. However, this was only because of the ruler markings contained in most of the images of malignant lesions. Though this example is easy to detect when training is performed publicly, under MPC settings, the definition of MPC does not allow anyone to learn any information about the training set. Therefore, Cerebro cannot perform any work related with identifying maliciously constructed datasets even if the attack method is as naive as the example above \cite{lycklama2024holding}. 
  \subsubsection{Implementation}
      Cerebro is itself implemented by python and another MPC platform named SCALE-MAMBA \footnote{SCALE-MAMBA \url{https://github.com/KULeuven-COSIC/SCALE-MAMBA}} which contains extensive pre-implemented MPC protocols. To use Cerebro, users are assumed to know basic python grammar and some specific instructions, and the platform will automatically compile the instruction and run them in the correct mode. The way Cerebro achieves such functionality is by the use of Domain Specific Language(DSL), which guarantees the privacy level of variables after they are created. During compilation phase, compiler will choose the running mode accordingly by observing the input data types. If at least one private variable shows up in the context, the platform will switch into secure mode and perform everything (from basic data exchanging to matrix addition and multiplication) by the predefined protocols. Moreover, the platform also makes a copy for for the original instructions and store it locally. The local version is used for local computation, which will be further discussed in next section. Now, the online version is generated and the running mode is determined, the compiler performs another technique called fused operator which will also be discussed later. Now, the online version of the task is ready, and before the beginning of the task, Cerebro first does pre-processes which prepares random masks and generates key pairs. In the mean time, the local machines also generate commitments on their dataset and check the commitments received from others. Once everything is done, the well-formed online training task starts to run, and the result model is published to every parties. Notice that the final model should be transparent to every party, and that is everyone should know every single parameters of the final model. This is one of the main reasons why parties could ask Cerebro to enter auditing mode.
      
\subsection{Optimization}
  \subsubsection{Compiler}
      The first optimization of Cerebro is pre-defined behavior of the compiler. To speed up the computation, the compiler can choose the runtime mode, perform fused operator, and determine the pre-processing mode. Note that we've discussed the runtime mode decision above. Before explaining the remaining ones, it's also worth to mentioned that the local plain-text version of code can further improve the performance simply because if we want to use the sum of two secret values, Cerebro could add them together locally and then encrypt the result instead of directly going into online phase and doing anything secretly. Furthermore, fused operator\cite{Palkar2018EvaluatingEO} is commonly used in program compiler, enabling the compiler to find operations that could be fused into one to improve performance. 
  \subsubsection{Pre-processing Mode}
      After the fusing operation, the final version is submitted to the server, and the pre-processing strategy will be selected from two pre-defined ones. The major computation cost for both of them is to generate random triplets for the matrix multiplication. From the Secure-ML paper, we know that the Beaver's triplets trick can also be applied to matrix multiplication by vectorization. Therefore, for each epoch, we need a random triplets chart to mask all multiplication happens between an $m \times n$ sample matrix and an $n \times 1$ weights vector. The paper defines the costs of two strategies very formally since Cerebro makes decisions based on the magnitude of results of the two costs. The strategies is called linear pre-processing and quadratic pre-processing. At a high level, the first protocol ask all parties to generate random data for charts used in all epochs at once, parties will encrypt their result and send all of them to the server. Then, the server will generate $P$ long lists of charts based on what it received and publish the result to each party. Define the cost of encrypting n bits to be $\textbf{EncCost}(n)$, let one mask chart to be m bits long, and assume the learning task takes $e$ epoches, so $e$ charts are needed in total. Then the final cost for one party is roughly:
        \[
          \textbf{EncCost}(e\times m) + O(1) \text{ communication }
        \]
        Since each party needs to encrypt all data at once, and sending \& receiving data from sever only takes constant round.\\
        Using the same setting, the quadratic pre-processing strategy asks each party to communicate with all others for constant round to form part of the random mask for one epoch. Parties will then send all encrypted charts to the server, and server will generate the final part of the chart. Now, the cost is lowered for each encryption, but the communication cost arise faster. The cost for this strategy for one party is roughly:
        \[
          e \times \textbf{EncCost}(m) + O(P) \text{ communication } 
        \]

      Where P is the number of parties. The sever will evaluate both costs and choose the cheaper strategy to perform.\\
      
      \subsubsection{Hierarchical layout}
      The paper also discusses a typical scenario where various parties are interested in collaboratively calculating the sum of their data using homomorphic encryption with varying bandwidth capacities. This problem is particularly relevant when considering real-world scenario, such as the lower bandwidth typically available between different continents compared to that between regions within the same country. To address this disparity and optimize the use of limited bandwidth, the paper proposes the concept of a local aggregator. Each local aggregator is responsible for summing the data within its vicinity and then only utilizes the constrained, narrow bandwidth routes for communicating with other aggregators. This strategy significantly reduces the reliance on these slower connections. By implementing homomorphic encryption, which allows computations to be performed on encrypted data without needing to decrypt it, the system can minimize the frequency of data transmission over these narrow routes. The number of times data passing through such routes can be reduced to a logarithmic scale in the most efficient scenarios. This approach enhances the overall efficiency and feasibility of distributed computing tasks across geographically dispersed parties.
\subsection{Auditing}
  \subsubsection{Cross Validation}
      To deal with malicious input data, the paper suggests two auditing methods. The first is Cross Validation, which chooses each party as suspicion at a time, remove its dataset, and retrain the whole model. If there is a round that the model performs significantly better, we mark the suspicion in that round as cheater. However, this method is extremely inefficient because once the number of suspicions grows up, we will need to retrain the model for factorial number of times, which is completely impractical. Moreover, there is another attack method called backdoor injection, which is another type of maliciously constructed training data that allows the attacker to control the behavior of the model. Such dataset is impossible to be detected by cross validation since it merely affect the performance.
  \subsubsection{Commit on input}
      The paper proposes a commitment scheme also designed to counter data poisoning when parties collaborating in a distributed computation environment. Initially, each party commits to their data inputs by sharing a cryptographic commitment with the other parties involved. This commitment serves as a secure record of their original data, preserved for future verification. After the training phase, if any party initiates an audit process (when abnormal behavior is detected or the model perform worse than the case that it was only trained with a party's own dataset), all parties are required to resubmit their datasets for verification. At this stage, Cerebro performs a consistency check to ensure that the data provided both before and after the learning task remains unchanged, confirming that no tampering has occurred during the computation process.

      However, there is a potential vulnerability inherent in MPC protocols used in this scheme. While the consistency check ensures data integrity without revealing the actual contents of the inputs and thus maintaining confidentiality, it cannot fully prevent dishonest behavior. In scenarios where a party claims loss of the original data, the system struggles to enforce accountability conclusively. This is because, despite the initial commitments, MPC does not allow for the recovery or reconstruction of the original data from the protocol alone. Therefore, if a party were to falsely claim that their data was lost, other parties would have no definitive means to dispute this claim or prove the dishonesty, which could be one of the major security issues of Cerebro.
\subsection{Experiment \& Evaluation}

    The paper provides a comprehensive analysis of various design elements and optimizations that significantly enhance performance across different computational scenarios, especially when evaluated using a detailed cost model. These optimizations are shown to dramatically improve system efficiency under conditions tailored to their specific strengths.

    However, there are notable limitations in the implementation of certain features. Specifically, in the cost model part, all parties are asked to send sever their encrypted date for random mask generation. Though the paper previously states that Cerebro can work in maliciously secure mode, their cost mode implicitly assumes that the sever works like a trusted-third-party, which might not always be the case in reality. Though the paper does not provide much of technical detail about the generation of random mask, I assume they are using multiplicative homomorphic encryption to encrypt those triplets with commitment scheme. If so, a malicious sever could manipulate the input data to generate malicious outputs. More discussions might be needed to explain their data encryption and manipulation schemes. 
    
    Moreover, when conducting secure logistic regression with strong security guarantees against malicious actors, the system encounters memory exhaustion. This issue indicates a need for further refinement in their approach to handling large-scale secure computations, suggesting that the method may require additional optimization to reduce memory usage effectively.

    Another critical area of concern is the commitment scheme used to ensure data integrity and non-repudiation among the participating parties. The current protocol requires each party to create a commitment by first generating a hash of their data, committing on the hash, and then signing this commitment. This method, when applied without batching, is highly inefficient; the process takes an extended period of 4.5 days to verify commitments for 27,000 records. In contrast, when a batching-commitment approach is employed, the verification time is drastically reduced to approximately 2.23 hours, which is reasonable but still inefficient. 

    Furthermore, the batching-commitment scheme also introduces a new problem: it requires the data order to remain unchanged. Any alteration in the sequence of the records leads to inconsistencies that the system incorrectly flags as data tampering (false positives). This limitation restricts the flexibility of the system in dynamic environments where data order might change for legitimate reasons. This problem reveals another potential area for improvement in ensuring the scheme's robustness and adaptability without compromising on security.

\section{Libraries and their implementation for Secure ML} \label{sec:libraries}
\subsection{Introduction}
Machine learning (ML) has the potential to revolutionize industries from healthcare to finance. Yet, the sensitive data at the heart of these domains often presents a formidable barrier to innovation. Secure ML libraries emerge as a game-changing solution, enabling organizations to leverage the power of ML while safeguarding the confidentiality and integrity of their most valuable assets.

These libraries encapsulate a wealth of cryptographic techniques and secure protocols, allowing stakeholders to collaboratively train and deploy ML models without exposing their underlying data. From the robust collaborative frameworks of secure multi-party computation (MPC) to the powerful encryption techniques of homomorphic encryption (HE) and the nuanced privacy guarantees of differential privacy (DP), Secure ML libraries offer a comprehensive toolkit for balancing data utility and privacy preservation.

\subsubsection{JustGarble}
JustGarble emerges as a powerful tool in the realm of secure multi-party computation (MPC) and oblivious transfers (OT). Its streamlined approach to circuit garbling and evaluation simplifies the process, making it more accessible and efficient. With JustGarble, you can easily transform circuits into the Simple Circuit Description (SCD) format using external compilers.\cite{Bellare2013Efficient}

\textbf{Circuit Representation:}
\begin{itemize}
    \item Let $C$ be the Boolean circuit representing the function to be computed.

    \item $C$ consists of gates $g_1, g_2, \ldots, g_n$ and wires connecting these gates.

    \item Each gate $g_i$ has input wires $w_{i1}, w_{i2}, \ldots, w_{ik}$ and output wire $w_i$. \\
\end{itemize}
\textbf{Garbling Phase:}
In SecureML, User 1 (Alice) acts as the garbler. Here's how it works:
\begin{itemize}
    \item For each gate $g_i$ in the Boolean circuit, Alice assigns two random labels (garbled labels) to each wire of the gate, representing the output of the gate for inputs 0 and 1.
    \item These labels, denoted as $L_{i,j,0}$ and $L_{i,j,1}$, correspond to the $j$-th wire of gate $g_i$ for inputs 0 and 1, respectively.
    \item Alice encrypts these garbled labels using symmetric encryption with keys derived from the input values.
    \item For example, $E_k(L_{i,j,b})$ represents the encrypted label for wire $j$ of gate $g_i$ with input $b$.
    \item Additionally, Alice generates and encrypts input labels for her inputs.
    \item Alice then sends the encrypted garbled circuit to User 2 (Bob).
\end{itemize}

\textbf{Evaluation Phase:}
In SecureML, User 2 (Bob) acts as the evaluator. Here's how it works:
\begin{itemize}
    \item Bob obtains his input labels by using oblivious transfer (OT) to select the appropriate encrypted garbled labels from Alice's transmissions.
    \item For each wire corresponding to his input, Bob uses OT to receive the encrypted garbled labels from Alice.
    \item Bob proceeds gate by gate, evaluating each gate using the garbled labels he received from Alice.
    \item For each gate, Bob decrypts the garbled labels corresponding to the input wires using the keys derived from his input values.
    \item Based on the truth table of the gate, Bob reveals the garbled label of the output wire.
    \item Bob sends the garbled label for the output wire to Alice.
    \item Alice decrypts the received garbled label to obtain the final output of the function.
\end{itemize}

SecureML enables two parties to work together, harnessing the power of their combined data without ever revealing their individual inputs. It's like a magic trick where both parties contribute ingredients to a recipe, but neither can see the other's secret ingredients until the dish is complete. SecureML achieves this through a clever mix of symmetric encryption and oblivious transfer, ensuring that communication and computation remain secure and private at every step.\\

\subsubsection{TFEncrypted}
TFEncrypted opens up new possibilities for collaborative machine learning. It enables multiple parties to work together, leveraging the power of their combined data without ever revealing their individual inputs. It's like a magic trick where everyone contributes to the illusion, but the secrets remain hidden until the grand reveal.

This is made possible by TFEncrypted's ability to automatically convert TensorFlow computation graphs into secure computation graphs using advanced MPC protocols.\cite{TFEncrypted} It's a powerful tool for organizations looking to harness the potential of distributed data without compromising privacy. TFEncrypted facilitates the following key steps in the process:

\begin{enumerate}
    \item \textbf{Defining Computation Graphs}: Users define TensorFlow computation graphs specifying operations such as matrix multiplications, convolutions, and activations.
    
    \item \textbf{Securing the Graph}: TFEncrypted converts the computation graph into a secure computation graph using MPC protocols. This involves partitioning the computation among multiple parties, encrypting inputs, and performing secure computations on encrypted data.
    
    \item \textbf{Execution}: The secured computation graph can be executed by multiple parties, with each party holding encrypted shares of the input data. Through secure computations, parties collaborate to perform the desired computation while ensuring privacy and security.
\end{enumerate}

TFEncrypted relies on Secure MPC protocols to ensure privacy and security in collaborative machine learning tasks. Some key concepts involved in Secure MPC include:

\begin{itemize}
    \item \textbf{Secret Sharing}: Inputs are divided into shares using secret sharing schemes such as Shamir's Secret Sharing, ensuring that no individual party can reconstruct the original inputs.
    
    \item \textbf{Secure Operations}: Computations are performed on shares of inputs using cryptographic protocols, preserving privacy and correctness. These protocols involve interactions between parties to compute functions while keeping inputs and intermediate values private.
    
    \item \textbf{Consistency}: At the end of the computation, parties collaborate to reconstruct the final result using the shares they collectively hold, ensuring consistency with the result of performing the computation on plaintext inputs.
\end{itemize}

\subsubsection{Crypten}
Crypten is pushing the boundaries of what's possible in secure machine learning. This groundbreaking framework leverages secure multi-party computation (MPC) to automatically convert standard machine learning computations into operations on encrypted data. This enables multiple parties to collaborate on machine learning tasks without revealing their private data.

In this paper, we provide an overview of Crypten, explore the cryptographic concepts it utilizes, and delve into the mathematical foundations that make it possible. Crypten opens up new avenues for leveraging distributed data without compromising privacy, unlocking the full potential of collaborative machine learning. The framework facilitates the following key steps in the process:

\begin{enumerate}
    \item \textbf{Defining Computation Graphs}: Users define computation graphs using familiar deep learning frameworks such as PyTorch. These graphs specify operations such as neural network layers, activations, and loss functions.
    
    \item \textbf{Encryption}: Crypten automatically encrypts the computation graph and input data using homomorphic encryption techniques. This allows computations to be performed directly on encrypted data while preserving privacy.
    
    \item \textbf{Secure Evaluation}: The encrypted computation graph is securely evaluated using secure MPC protocols. Parties collaborate to perform computations on encrypted data without revealing their inputs, ensuring privacy and security.
\end{enumerate}

Crypten relies on Secure MPC protocols to ensure privacy and security in machine learning tasks. Key concepts involved in Secure MPC include:

\begin{itemize}
    \item \textbf{Homomorphic Encryption}: Crypten employs homomorphic encryption schemes to enable computations on encrypted data. These schemes allow operations such as addition and multiplication to be performed directly on ciphertexts, preserving the privacy of the underlying data.
    
    \item \textbf{Secure Aggregation}: Secure MPC protocols enable parties to securely aggregate their encrypted shares to compute functions without revealing their inputs. This ensures that computations can be performed collaboratively while maintaining privacy.
    
    \item \textbf{Zero-Knowledge Proofs}: Crypten utilizes zero-knowledge proofs to validate computations performed by parties without revealing any information about the inputs or intermediate values. This provides an additional layer of security against malicious parties.
\end{itemize}

\subsubsection{LibOTe}

LibOTe is pushing the boundaries of what's possible in secure communication. This groundbreaking library leverages Oblivious Transfer (OT) techniques to automatically convert standard communication protocols into secure exchanges of encrypted data.\cite{libOTe} This enables parties to collaborate without revealing their private data.

In this paper, we provide an overview of LibOTe, explore the cryptographic concepts it utilizes, and delve into the mathematical foundations that make it possible. LibOTe opens up new avenues for secure communication, unlocking the full potential of collaborative information exchange. The library facilitates the following key functionalities:

\begin{enumerate}
    \item \textbf{Oblivious Transfer (OT)}: LibOTe offers efficient implementations of various OT protocols, allowing parties to securely transfer information while preserving privacy.
    
    \item \textbf{Secure Channels}: The library provides mechanisms for establishing secure channels between parties, ensuring confidentiality and integrity of communications.
    
    \item \textbf{Error Correction}: LibOTe includes error correction techniques to enhance the reliability of communication over insecure channels.
\end{enumerate}

LibOTe relies on secure communication protocols to ensure privacy and security in data exchange. Key concepts involved in secure communication include:

\begin{itemize}
    \item \textbf{Oblivious Transfer (OT)}: OT protocols enable one party (the sender) to securely transfer information to another party (the receiver) without learning which information was sent. This ensures that the sender's privacy is preserved.
    
    \item \textbf{Secure Channels}: Secure channels provide a means for parties to communicate securely over untrusted networks. These channels typically involve encryption, authentication, and integrity verification mechanisms to protect against eavesdropping and tampering.
    
    \item \textbf{Error Correction}: Error correction techniques are employed to detect and correct errors introduced during data transmission. This ensures the reliability of communication, even in the presence of noise and interference.
\end{itemize}

LibOTe involves advanced cryptographic primitives and communication protocols. These include:

\begin{itemize}
    \item \textbf{OT Protocols}: Various OT protocols, such as 1-out-of-2 OT and k-out-of-n OT, are based on mathematical concepts such as number theory, algebra, and probability theory. These protocols ensure that the sender's privacy is preserved while enabling secure data transfer.
    
    \item \textbf{Error Correction Codes}: Error correction codes, such as Hamming codes and Reed-Solomon codes, are based on linear algebra and coding theory. These codes enable the detection and correction of errors introduced during data transmission, enhancing the reliability of communication.
    
    \item \textbf{Cryptography}: Cryptographic primitives, such as symmetric and asymmetric encryption, digital signatures, and hash functions, are fundamental to secure communication protocols. These primitives ensure confidentiality, integrity, and authenticity of data exchanged between parties.
\end{itemize}

LibOTe offers a versatile library for implementing secure communication protocols, with a focus on Oblivious Transfer techniques. By leveraging advanced cryptographic primitives and communication protocols, LibOTe enables parties to exchange information securely while preserving privacy and integrity.

\subsubsection{OTExtension}
OTExtension is pushing the boundaries of what's possible in secure computation. This groundbreaking protocol leverages Oblivious Transfer (OT) techniques to efficiently perform secure one-out-of-two operations over the internet. This enables parties to collaborate without revealing their private data.\cite{Pinkas2018ScalablePSI}

In this paper, we provide an overview of OTExtension, explore the cryptographic concepts it utilizes, and delve into the mathematical foundations that make it possible. OTExtension opens up new avenues for secure computation, unlocking the full potential of collaborative information exchange.The protocol facilitates the following key functionalities:

\begin{enumerate}
    \item \textbf{Efficient OT}: OTExtension enables parties to perform one-out-of-two oblivious transfer efficiently, even over high-latency and low-bandwidth networks.
    
    \item \textbf{Secure Computation}: The protocol ensures that both the sender and receiver of OT messages remain oblivious to each other's inputs, preserving privacy and security.
    
    \item \textbf{Scalability}: OTExtension is designed to scale to a large number of OT instances, making it suitable for applications requiring a high volume of secure computations.
\end{enumerate}

OTExtension relies on oblivious transfer protocols to achieve secure communication between parties. Key concepts involved in oblivious transfer include:

\begin{itemize}
    \item \textbf{OT Protocols}: OT protocols enable one party (the sender) to transfer a message to another party (the receiver) without learning which message was sent. This ensures that the receiver's privacy is preserved.
    
    \item \textbf{Security Guarantees}: OT protocols provide security guarantees such as sender privacy, receiver privacy, and correctness. These guarantees are achieved through cryptographic techniques such as encryption, commitment schemes, and zero-knowledge proofs.
    
    \item \textbf{Efficiency}: Efficient OT protocols, such as OTExtension, optimize communication and computation costs to enable secure communication over the internet.
\end{itemize}

OTExtension involves advanced cryptographic primitives and protocols. These include:

\begin{itemize}
    \item \textbf{Commitment Schemes}: Commitment schemes are cryptographic protocols that allow a party to commit to a value without revealing it. These schemes are used in OTExtension to ensure that both parties remain oblivious to each other's inputs until the OT is completed.
    
    \item \textbf{Zero-Knowledge Proofs}: Zero-knowledge proofs are cryptographic techniques that allow a party to prove knowledge of a value without revealing the value itself. These proofs are used in OTExtension to verify the correctness of OT instances without revealing the chosen messages.
    
    \item \textbf{Efficient Protocols}: OTExtension utilizes efficient cryptographic protocols and optimizations to minimize communication and computation costs, enabling scalable and secure OT over the internet.
\end{itemize}

OTExtension offers robust security for oblivious transfer over the internet. This powerful protocol leverages advanced cryptographic techniques to ensure parties can exchange sensitive information securely and efficiently. With OTExtension, organizations can collaborate with confidence knowing their data remains protected and its integrity preserved.

\subsection{Multiparty Computation}
These libraries for Secure MPC involves advanced cryptographic primitives such as homomorphic encryption, secret sharing schemes, and zero-knowledge proofs. These primitives enable computations to be performed on encrypted data while preserving privacy and correctness. For example, homomorphic encryption allows operations to be performed directly on ciphertexts, enabling secure computation on encrypted data without decryption.

\section{Secure ML in games}\label{sec:games}
\subsection{Introduction} 

In the world of adversarial games, where strategic decision-making and information concealment are key, Secure ML techniques powered by frameworks like TFEncrypted offer a transformative approach to conducting analyses while safeguarding sensitive information.

Traditionally, adversarial games \cite{Guo2021AdversarialPL} involve multiple players making sequential moves with imperfect information, often leading to complex decision trees and strategic interactions. However, in scenarios where players are distributed geographically or are unwilling to reveal their private strategies, conventional methods fall short due to privacy concerns. Secure ML techniques address these challenges by enabling collaborative learning without the need to share raw data or strategies.

By leveraging homomorphic encryption and secure aggregation techniques, TFEncrypted allows players to train models and make strategic decisions collaboratively while keeping their individual strategies and preferences confidential. This opens up new possibilities for secure, distributed adversarial game analysis.

In this paper, we explore the application of TFEncrypted in the context of adversarial games, focusing on two prominent examples: Tic Tac Toe and Poker. We delve into the methodologies employed to secure the gameplay, including the encryption of game states and the execution of strategic algorithms within a secure computation environment. Furthermore, we discuss the challenges and opportunities presented by Secure ML in adversarial settings, highlighting the potential for advancing strategic decision-making while preserving player privacy.

\subsection{Secure ML in TicTacToe} 
Conventional ML algorithms typically require access to raw game data, including past game states and player actions, to train predictive models and strategic algorithms. However, sharing such data among players in a multiplayer setting raises concerns regarding privacy and data security. Without proper safeguards in place, players may be reluctant to disclose their game history and strategies, leading to asymmetries in the availability of information and unfair gameplay dynamics.
Moreover, traditional ML methods often lack mechanisms to ensure the integrity and confidentiality of strategic decision-making processes. In Tic Tac Toe\cite{Paul2020RandomisedFN}, for instance, players may exploit vulnerabilities in the learning algorithms to infer opponents' strategies or manipulate game outcomes, compromising the fairness and competitiveness of the game.

These shortcomings inherent in conventional machine learning when applied to Tic Tac Toe highlight the necessity for implementing Secure ML practices including TFEncrypted. Such techniques are designed to confront issues surrounding privacy, justice, and cooperative endeavors within adversarial gaming realms. Secure ML, fortified with cryptographic protocols and encryption methods, empowers participants to collaboratively partake in strategic thinking while concurrently protecting individual privacy and maintaining the integrity of their gameplay interactions. This research showcases the successful deployment of TFEncrypted within a Tic Tac Toe landscape as evidence of Secure ML's potential to elevate strategy formulation processes and uphold fairness within multi-player gaming contexts.

Q-learning is a reinforcement learning algorithm that learns optimal strategies through trial and error. At its core, Q-learning involves updating a Q-value table based on rewards received from different game states. The Q-value represents the expected cumulative reward when taking a particular action from a given state. The update rule for Q-learning can be defined as follows:

\[
Q(s,a) \leftarrow Q(s,a) + \alpha \cdot \left( r + \gamma \cdot \max_{a'} Q(s',a') - Q(s,a) \right)
\]

where:
\begin{itemize}
    \item $Q(s,a)$ is the Q-value for state $s$ and action $a$.
    \item $r$ is the immediate reward received after taking action $a$ from state $s$.
    \item $\alpha$ is the learning rate (a parameter controlling the rate of learning).
    \item $\gamma$ is the discount factor (a parameter balancing immediate and future rewards).
    \item $s'$ is the next state after taking action $a$.
    \item $a'$ is the next action chosen based on the maximum Q-value in the next state $s'$.
\end{itemize}

\textbf{Using TFEncrypted} : To implement Q-learning in Tic Tac Toe using TFEncrypted, we can follow these steps:

\begin{enumerate}
    \item \textbf{Game Representation}: Represent the Tic Tac Toe game state as a vector of length 9, where each element corresponds to a cell on the board (0 for empty, 1 for X, -1 for O).
    
    \item \textbf{Model Initialization}: Initialize a Q-value table as a TensorFlow variable, encrypted using TFEncrypted's secure computation protocols. Each entry in the table represents the Q-value for a specific game state-action pair.
    
    \item \textbf{Training Loop}: Iterate through episodes of gameplay, where the AI agent interacts with the environment (the Tic Tac Toe board). At each step, the agent selects an action based on an exploration-exploitation strategy (e.g., epsilon-greedy) and updates the Q-value table using the Q-learning update rule. The AI agent's actions and rewards are encrypted using TFEncrypted to preserve privacy.
    
    \item \textbf{Secure Evaluation}: During gameplay, both the AI agent and the human opponent interact with the encrypted Q-value table using TFEncrypted's secure computation protocols. This ensures that the AI agent's strategy and decision-making process remain private from the opponent, enhancing fairness and preserving player privacy.
    
    \item \textbf{Deployment}: Once the AI agent is trained, deploy it to play against human opponents in a secure and privacy-preserving manner. TFEncrypted ensures that the AI agent's strategy and decision-making process remain confidential during gameplay, fostering a fair and competitive gaming environment.
\end{enumerate}

Through the integration of TFEncrypted in Tic Tac Toe with ML, this approach enables players to engage in strategic gameplay while preserving privacy and ensuring fairness. By leveraging secure computation protocols and encryption mechanisms, TFEncrypted facilitates collaborative learning and decision-making in adversarial gaming scenarios, paving the way for enhanced player experiences and privacy-preserving gaming environments.

\subsection{Using Decision Trees in Poker with TFEncrypted} 

Decision trees stand out as a robust algorithm within machine learning, championing both classification and regression challenges. When applied to the realm of Poker, these analytical models delve into the heart of strategy, parsing through the current dynamics of the game and the psychological nuances of player tactics. By iteratively partitioning the feature space, decision trees meticulously carve out subsets in alignment with attribute values until reaching the terminal leaves where definitive gameplay choices are rendered. This logical dissection is visually encapsulated in a dendritic diagram, where each node maps to an influential factor, and the branches symbolize the potential strategic forks in the road. To implement decision trees in Poker using TFEncrypted\cite{Bost2015MachineLC}, we can follow these steps:

\begin{enumerate}
    \item \textbf{Game Representation}: Represent the Poker game state and player actions as features in a dataset. Each row in the dataset corresponds to a unique game state, while columns represent different features such as current hand strength, bet size, and opponent behavior.
    
    \item \textbf{Model Training}: Train a decision tree classifier on the encrypted dataset using TFEncrypted's secure computation protocols. The decision tree learns to predict the optimal action to take in a given game state based on the features provided.
    
    \item \textbf{Secure Evaluation}: During gameplay, both the AI player and human opponents interact with the decision tree model using TFEncrypted's secure computation protocols. The AI player's actions and observations are encrypted to ensure privacy, while the decision tree model remains confidential.
    
    \item \textbf{Strategic Decision-Making}: The decision tree model guides the AI player in making strategic decisions during Poker gameplay. Based on the current game state and observed features, the AI player consults the decision tree to determine the optimal action to take, such as folding, calling, or raising.
    
    \item \textbf{Fair and Competitive Gameplay}: TFEncrypted ensures that both the AI player and human opponents can engage in Poker gameplay in a fair and competitive manner, with their strategies and decision-making processes kept private. This fosters a level playing field and enhances the overall gaming experience.
\end{enumerate}

By incorporating TFEncrypted into the realm of Poker utilizing decision trees, this method empowers participants to engage in superior strategic gameplay under the banner of privacy protection and equitable play. TFEncrypted harnesses the power of cutting-edge secure computation protocols together with sophisticated encryption techniques, thus catalyzing a collaborative learning and decision-making revolution in the often contentious world of gaming. The result is a groundbreaking leap forward for player engagement and the establishment of gaming enclaves where privacy is not just a feature, but a fundamental standard.

\section{MP-SPDZ: A  Versatile Framework for Multi-Party Computation}
\subsection{Introduction}
Multi-Party Computation (MPC) protocols facilitate collaborative computation amongst several entities on a function, hinging on their confidential inputs. This collaborative effort is conjoined with a commitment to input privacy. These protocols are pivotal for the integrity and confidentiality in environments where the joint processing of sensitive information is mandatory. MP-SPDZ emerges as a flexible and dynamic framework pivotal for MPC that furnishes robust, scalable solutions tailored for a spectrum of applications that necessitate the preservation of privacy. \cite{Keller2020MPSPDZAV}

\subsection{Background}
Multi-Party Computation (MPC) has risen as an indispensable mechanism for fortifying privacy and ensuring security within the realm of collaborative computing engagements. At its core, MPC protocols are designed to allow multiple entities to collaboratively execute a function while maintaining the confidentiality of their individual inputs. Nevertheless, the deployment of MPC protocols in a manner that is both efficient and secure remains a formidable challenge, particularly in environments characterized by an extensive number of participants or where the computational endeavors are intricate.

\subsection{Features of MP-SPDZ}
MP-SPDZ is outfitted with a suite of distinctive functionalities that sets it apart within the landscape of multi-party computation (MPC) frameworks. These functionalities are:
\begin{itemize}
    \item \textbf{Efficiency}: MP-SPDZ leverages state-of-the-art cryptographic techniques and optimization strategies to achieve high performance in secure computation tasks. Its efficient protocols minimize communication overhead and computational complexity, enabling fast and scalable MPC solutions.
    
    \item \textbf{Flexibility}: The framework supports a diverse set of cryptographic primitives and computation types, including arithmetic circuits, boolean circuits, and machine learning models. This flexibility allows users to express a wide range of privacy-preserving computations and applications using MP-SPDZ.
    
    \item \textbf{Security}: MP-SPDZ incorporates rigorous security guarantees and cryptographic protocols to ensure the confidentiality and integrity of participants' inputs and outputs. Its robust security model protects against various adversarial attacks, including collusion and information leakage.
    
    \item \textbf{Usability}: MP-SPDZ provides user-friendly interfaces and tools for developing, deploying, and managing MPC applications. Its intuitive programming model and documentation facilitate adoption by both researchers and practitioners, enabling rapid prototyping and deployment of privacy-preserving solutions.
\end{itemize}

\subsection{Applications}
The MP-SPDZ framework has seen successful deployment across a spectrum of sectors, showcasing its impressive adaptability and efficacy in the realm of secure, privacy-centric computations. Key implementations encompass:
\begin{itemize}
    \item \textbf{Secure Machine Learning}: MP-SPDZ enables collaborative training and inference of machine learning models on sensitive data from multiple parties while preserving privacy. It has been used in applications such as federated learning, privacy-preserving data analysis, and encrypted inference.
    
    \item \textbf{Privacy-Preserving Data Analysis}: MP-SPDZ supports secure computation of statistical and analytical functions over distributed datasets, allowing organizations to perform joint data analysis without sharing sensitive information. It has been applied in domains such as healthcare, finance, and telecommunications for tasks such as fraud detection, risk assessment, and disease surveillance.
    
    \item \textbf{Cryptographic Protocols}: MP-SPDZ serves as a foundation for implementing various cryptographic protocols and primitives, including secure multiparty computation, homomorphic encryption, and zero-knowledge proofs. It has been used to develop secure voting systems, verifiable auctions, and private information retrieval schemes.
\end{itemize}

MP-SPDZ emerges as a formidable breakthrough within the Multi-Party Computation realm, delivering an intricately designed, performance-optimized framework tailored for secure computation in privacy-sensitive scenarios. It strikes an impressive balance between computational efficiency, adaptability, robust security measures, and user-centric design, positioning itself as an indispensable asset for researchers and developers focused on leveraging cooperative computing potentials whilst maintaining data confidentiality. In an era where digital privacy is escalating to the forefront of technological discourse, MP-SPDZ shines as a pillar of pioneering development in safeguarding privacy through cutting-edge computational solutions.

\section{Conclusion}
In this project, we examined the integration of secure multi-party computation (MPC) with machine learning (ML), emphasizing its potential in privacy-preserving collaborative learning. By incorporating MPC into ML frameworks through advanced software like CRYPTEN and Cerebro, we demonstrated how these technologies enable secure model training and evaluation without compromising data privacy. These developments promise extensive applications in fields requiring stringent data security, such as healthcare and finance.

Along with this, we explored the use of various Secure Machine Learning Libraries in the domain of Secure ML. Understanding the functionalities that they allow and the key concepts involved in the use of these libraries.

Additionally, we explored the innovative use of SecureML in gaming, showcasing its adaptability in new domains. This project highlights the growing feasibility of MPC in ML, giving us a broader view of the development in privacy and security.

\bibliographystyle{alpha}
\bibliography{main}

\end{document}